%% file: main.tex
\documentclass[sigconf, screen]{acmart}

\usepackage{graphicx, centernot, amsmath, bbm, booktabs, multirow, array, capt-of, varwidth}

\usepackage[suppress]{color-edits}

\addauthor{sw}{blue}
\addauthor{kh}{magenta}
\addauthor{lg}{orange}
\addauthor{ac}{red}

\newcommand{\bigCI}{\mathrel{\text{\scalebox{1.07}{$\perp\mkern-10mu\perp$}}}}

\newcommand{\CI}{\mathrel{\perp\mspace{-10mu}\perp}}
\newcommand{\nCI}{\centernot{\CI}}
\newcolumntype{V}[1]{>{\begin{varwidth}[t]{#1}}l<{\end{varwidth}}}

\AtBeginDocument{%
  \providecommand\BibTeX{{%
    \normalfont B\kern-0.5em{\scshape i\kern-0.25em b}\kern-0.8em\TeX}}}

\setcopyright{rightsretained}

\copyrightyear{2023}
\acmYear{2023}
\acmConference[FAccT '23]{2023 ACM Conference on Fairness, Accountability, and Transparency}{June 12--15, 2023}{Chicago, IL, USA}
\acmBooktitle{2023 ACM Conference on Fairness, Accountability, and Transparency (FAccT '23), June 12--15, 2023, Chicago, IL, USA}
\acmDOI{10.1145/3593013.3594036}
\acmISBN{979-8-4007-0192-4/23/06}
\begin{document}

\title{Ground(less) Truth: A Causal Framework for Proxy Labels in Human-Algorithm Decision-Making}

\author{Luke Guerdan}
\email{lguerdan@cs.cmu.edu}
\affiliation{%
  \institution{Carnegie Mellon University}
  \city{Pittsburgh}
  \state{PA}
  \country{USA}
}

\author{Amanda Coston}
\email{acoston@cs.cmu.edu}
\affiliation{%
  \institution{Carnegie Mellon University}
  \city{Pittsburgh}
  \state{PA}
  \country{USA}
}

\author{Zhiwei Steven Wu}
\email{zstevenwu@cmu.edu}
\affiliation{%
  \institution{Carnegie Mellon University}
  \city{Pittsburgh}
  \state{PA}
  \country{USA}
}

\author{Kenneth Holstein}
\email{kjholste@cs.cmu.edu}
\affiliation{%
  \institution{Carnegie Mellon University}
  \city{Pittsburgh}
  \state{PA}
  \country{USA}
}

\input{Sections/0_abstract}

\maketitle

\input{Sections/1_introduction}

\input{Sections/2_related_work}

\input{Sections/3_framework}

\input{Sections/4_modeling}

\input{Sections/5_experiments}
\input{Sections/6_discussion}

\bibliographystyle{ACM-Reference-Format}
\bibliography{refs}

\input{Sections/7_appendix}

\end{document}

%% file: Sections/0_abstract.tex
\begin{abstract}
A growing literature on human-AI decision-making investigates strategies for combining human judgment with statistical models to improve decision-making. Research in this area often evaluates proposed improvements to models, interfaces, or workflows by demonstrating improved predictive performance on ``ground truth’’ labels. However, this practice overlooks a key difference between human judgments and model predictions. Whereas humans commonly reason about broader phenomena of interest in a decision – including latent constructs that are not directly observable, such as disease status, the ``toxicity'' of online comments, or future ``job performance'' – predictive models target \textit{proxy} labels that are readily available in existing datasets. Predictive models' reliance on simplistic proxies for these nuanced phenomena makes them vulnerable to various sources of statistical bias. In this paper, we identify five sources of \textit{target variable bias} that can impact the validity of proxy labels in human-AI decision-making tasks. We develop a causal framework to disentangle the relationship between each bias and clarify which are of concern in specific human-AI decision-making tasks. We demonstrate how our framework can be used to articulate implicit assumptions made in prior modeling work, and we recommend evaluation strategies for verifying whether these assumptions hold in practice. We then leverage our framework to re-examine the designs of prior human subjects experiments that investigate human-AI decision-making, finding that only a small fraction of studies examine factors related to target variable bias. We conclude by discussing opportunities to better address target variable bias in future research. 
\end{abstract}


\ccsdesc[500]{Human-centered computing~Human computer interaction (HCI)}
\ccsdesc[500]{Human-centered computing~User studies}
\ccsdesc[500]{Computing methodologies~Machine learning}

\keywords{algorithmic decision support, measurement, validity, causal diagrams, label bias, human-AI decision-making}

%% file: Sections/1_introduction.tex
\section{Introduction}
A growing body of research \khedit{aims to combine}\khdelete{combines} predictive machine learning \khedit{(ML)} models with human judgment to improve decision-making processes. \lgdelete{This work often pursues the goal of \textit{complementary performance}: configurations of humans and models that yield higher-quality decisions than either would make in isolation.}\khedit{In the machine learning community, researchers have proposed improvements to \textit{ML models} to better}\khdelete{In the machine learning community, model-level improvements have been proposed to} address gaps in human judgment (e.g., \citep{madras2018predict, wilder2020learning, tan2018investigating, hilgard2021learning}). In the human-computer interaction community, behavioral interventions have been developed to help \textit{humans} better incorporate model outputs into their decision-making (e.g., \citep{bansal2021does, buccinca2021trust, liu2021understanding, buccinca2020proxy, carton2020feature,bussone2015role, lai2021towards}). \lgedit{However, current evaluations of \khedit{both} model-level \khedit{and human}\khdelete{improvements and} behavioral interventions \khedit{typically}\khdelete{frequently} assess the quality of human decisions, algorithm\khedit{ic} predictions, and hybrid combinations of the two by comparing their accuracy on ``ground-truth’’ labels \khedit{that are} readily available in \khedit{existing} data. \khedit{This}\khdelete{Yet this} practice assumes that the labels targeted by predictive models serve as a reliable measure of the underlying goals and objectives of human decision-makers.} \lgdelete{ \textbf{However, current evaluations of human-AI decision-making make the key assumption that label targeted by a predictive model adequately reflects the goals of human decision-makers}. Decision quality is frequently operationalized by predictive performance, measured via accuracy, AU-ROC, or similar statistical measures computed with respect to \textit{``ground truth”} labels that are readily available in existing data. The relative performance of human judgment, model predictions, and hybrid combinations of the two are then ranked according to these metrics. Such comparisons of human and model performance are only valid insofar as the \textit{``ground truth"} label targeted by the model reflects the underlying objectives of human decision-makers.}

\lgdelete{In real-world human-AI decision-making settings}\khedit{Yet in}\lgedit{\khdelete{In} real-world deployments of algorithmic decision support (ADS) tools}, labels are often \lgedit{imperfect proxies} for the target outcomes \lgdelete{considered by decision-makers}\lgedit{of interest to human experts}. While making decisions, \lgdelete{humans frequently consider latent constructs such as the ``toxicity’' of online comments, ``cardiovascular disease risk’’ of a patient, or future ``job performance’’ of a candidate.} \lgedit{content moderators frequently assess the ``toxicity’' of online comments \citep{halfaker2020ores} while physicians often consider the ``cardiovascular disease risk’’ of patients \citep{anderson1991cardiovascular}. ``Toxicity’' and ``cardiovascular disease risk’’ are examples of latent constructs which are unobserved in data.} Because observed labels (e.g., toxicity annotations and diagnostic test results) serve as indirect measurements of these phenomena \citep{jacobs2021measurement}, they can be subject to \textit{measurement error}. Additionally, humans often select among multiple possible actions (e.g., medical treatments, social welfare interventions) in hopes of improving a downstream outcome of interest. Because an outcome is only observed for the selected action, labeled data does not contain the counterfactual outcome that \textit{would} occur had a different option been chosen instead. This introduces a set of additional challenges, including selective labels \citep{lakkaraju2017selective}, intervention effects \citep{coston2020counterfactual}, and selection bias \citep{rambachan2019bias}, which interact with measurement error in nuanced ways depending on the nature of the \khedit{specific} decision-support task. 

We refer to this collection of challenges – which can be characterized as sources of statistical bias impacting labels – as \textit{target variable bias} (TVB).\footnote{The term Target Variable Bias was introduced in \citep{chouldechova2018case, fogliato2020fairness}. We use this as an umbrella term describing sources of statistical bias known to impact proxy labels in decision support tasks. }  \lgedit{Following common terminology in statistics, we use the term ``bias'' to describe systematic differences between the target outcome of interest to human experts and its imperfect operationalization in available data. Thus, while TVB describes a broad conceptual difference between outcomes of interest and their observed proxies, this difference can be formally studied under existing statistical frameworks.} 

Target variable bias has been widely documented in real-world deployments of algorithmic systems \citep{kleinberg2018human, fogliato2021validity, bao2021s, butcher2022racial, kawakami2022improving, cheng2022disparities, obermeyer2019dissecting, mullainathan2017does, mullainathan2019machine, chalfin2016productivity}. Predictive models impacted by target variable bias have contributed to unwarranted firing of teachers \citep{chalfin2016productivity}, perpetuated \lgdelete{historical} disparities in access to medical resources \citep{obermeyer2019dissecting}, and raised concerns among social workers investigating allegations of child abuse and neglect \citep{kawakami2022improving, cheng2022disparities}. Surprisingly, existing modeling efforts and human subjects experiments in the human-AI decision-making literature have largely overlooked this challenge. \textbf{Left unaddressed, this disconnect could undermine the ultimate goal of human-AI decision-making research: to develop algorithmic systems that meaningfully improve decision-making in real-world contexts.} 

\lgedit{Therefore,} in this work, we bridge the \lgedit{divide} between challenges encountered in real-world deployments of predictive models and current human-AI decision-making research practices by (i) raising awareness of target variable bias, (ii) identifying \lgdelete{blind spots}\lgedit{gaps} in previously published modeling approaches and human subjects experiments, and (iii) providing guidelines for improved research practices going forward. \lgedit{In particular, we develop a \textit{causal framework} which identifies the sources and implications of target variable bias in human-AI decision-making by examining the data generating process which gives rise to predictive model training datasets. Our framework enables us to distill ADS tasks studied in prior literature into their underlying structural components, and identify which sources of TVB (e.g., measurement error, intervention effects, selective labels) are of concern in a specific task.} \lgdelete{Based on an examination of numerous real-world deployments of predictive models, we construct a causal framework that captures possible data generating processes in human-AI decision-making. Our framework distills high-level structural differences between various real-world decision support tasks and disentangles which sources of bias are of concern in each setting. We map our framework to existing terminology and methods used to characterize target variable modeling assumptions across multiple disciplines.} Using our framework, we identify opportunities to better\khdelete{-} address target variable bias through two lines of human-AI decision-making research:

\begin{itemize}
\item \textbf{Model development}. We develop a \textit{measurement} and \textit{prediction} decomposition that articulates target variable modeling assumptions. We use our decomposition to create a taxonomy of model-level improvements proposed in previous literature. We also propose a set of recommended measurement model evaluation strategies. 

\item \textbf{Experimental human subjects studies}. We \lgedit{use our framework to} re-examine the design of prior human subjects experiments studying human-AI decision-making. Our analysis identifies systematic \lgedit{gaps}\lgdelete{blind-spots} in our current understanding of human-AI decision-making due to target variable bias.
\end{itemize}

%% file: Sections/2_related_work.tex
\section{Related work}

We begin by introducing the body of human-AI decision-making research our framework is designed to inform. We then summarize modeling challenges and broader validity concerns that draw current research practices (i.e., modeling assumptions, experimental study designs, and measures of decision quality) into question. 

\subsection{Human-AI decision-making}

Recent machine learning research proposes techniques designed to complement the limitations of human judgement. Drawing from a long line of work showing that actuarial risk assessments can outperform expert judgement in many prediction tasks \citep{grove2000clinical, dawes1989clinical}, methods have been proposed that learn to complement humans by adaptively routing decision instances \citep{madras2018predict, gao2021human}, leveraging heterogeneity in human and machine decision performance \citep{tan2018investigating, wilder2020learning, donahue2022human, charusaie2022sample}, leveraging consistency in expert decisions \citep{de2021leveraging}, and adapting to \citep{hilgard2021learning} and training \citep{mozannar2022teaching} human mental representations of model outputs. Yet these techniques operate on a set of simplifying assumptions about the world, which may or may not hold in a given deployment context. We provide a framework for articulating modeling assumptions, and show that many common assumptions made by prior work involving proxy labels are unlikely to hold in practice. Recent research has also studied opportunities for human-AI complementary in algorithm-assisted \textit{human} decision-making \citep{lai2019human, green2019principles, buccinca2021trust, gajos2022people, cai2021onboarding, lai2020chicago}. This work investigates the potential for tools such as training protocols \citep{cai2021onboarding, lai2020chicago, cai2019hello}, explanations \citep{liu2021understanding, bussone2015role}, and other behavioral interventions \citep{buccinca2021trust, gajos2022people}, to improve how humans make use of model outputs. While many online experimental studies have \lgdelete{been}focused on interventions to improve predictive performance, little work to date has experimentally studied other key factors that are present in real-world deployment contexts, such as asymmetric access to information \citep{holstein2023toward, hemmer2022effect}, measurement error \citep{gordon2022jury}, and omitted payoffs \citep{green2021algorithmic}.

\subsection{Modeling challenges in algorithmic decision support}\label{subsec:modeling_challenges}

Prior work has surfaced a litany of challenges impacting predictive models designed for algorithmic decision support (ADS), including unobservables \citep{kleinberg2018human}, selective labels \citep{lakkaraju2017selective}, selection bias \citep{singh2021fairness, de2018learning}, and intervention effects \citep{coston2020counterfactual}. Additional work has examined the quality of proxy labels in decision support tasks. For example, \citet{obermeyer2019dissecting} surfaced \textit{``label choice bias''}, in which racial disparities in access to health resources were introduced by poor label selection decisions. \textit{``Omitted payoffs bias''} describes factors of interest to humans that are incompletely reflected by predictive models targeting available labels \citep{de2021leveraging, chalfin2016productivity, kleinberg2018human}. While this bias describes challenges specific to prediction (e.g., model unobservables, measurement error \citep{kleinberg2018human}), this term also applies when humans care about a broader set of decision-making factors beyond predictive risk \citep{chalfin2016productivity, green2019principles}. In this work, we use the lens of measurement \lgedit{and validity} to examine systematic differences between target outcomes of interest to humans and proxy labels observed in data \citep{jacobs2021measurement}. In adopting this lens, we draw upon a rich set of existing knowledge and methodologies from adjacent disciplines (e.g., psychology, political science, sociology) designed to evaluate how latent phenomena of interest to humans are quantified in data \citep{roberts1985measurement}.

\subsection{Measurement and validity in algorithmic systems}

Recent work has raised broader concerns regarding whether algorithmic systems successfully achieve their purported function \citep{coston2022validity, jacobs2021measurement,bao2021s}. Synthesizing concepts from measurement theory in the quantitative social sciences, \citet{jacobs2021measurement} argue that \textit{``algorithmic fairness''} is a latent construct that is imperfectly operationalized by statistical fairness measures. \citet{bao2021s} examine statistical biases present in criminal justice datasets (e.g., ProPublica's COMPAS Dataset \citep{angwin2016machine}) used in fairness benchmarks of algorithmic Risk Assessment Instruments (RAIs). This analysis identifies several biases in the outcome variable $Y$ target by models, which we further characterize in this work. \citet{coston2022validity} highlight validity concerns impacting RAIs, including many discussed in $\S$ \ref{subsec:modeling_challenges}. Recent work has also surfaced validity issues in content moderation \citep{gordon2022jury} and recommender systems \citep{milli2021optimizing, stray2022building}. 

Despite this growing awareness, we currently lack a holistic understanding of validity threats to prediction targets in human-AI decision-making. Addressing this gap is critical for preventing algorithmic harms in real-world deployment contexts. Therefore, in this work, we use causal diagrams to examine the relationship between measurement error and additional modeling challenges (i.e., $\S$ \ref{subsec:modeling_challenges}) that can impact the validity of prediction targets in real-world decision support settings. To our knowledge, our work offers the first holistic examination of how measurement error, unobservables, selection bias, intervention effects\lgedit{, and confounding} interact to impact target variable validity in real-world ADS deployments.

%% file: Sections/3_framework.tex
\section{Framework}\label{sec:framework}

We now describe our framework scope ($\S$ \ref{sec:scope}) and development process ($\S$ \ref{sec:process}) before introducing our causal diagram ($\S$ \ref{sec:causal_diagram}). We then use our framework to \lgedit{map algorithmic decision support tasks to relevant sources of target variable bias ($\S$ \ref{sec:tasks_challenges})}. \lgdelete{create a taxonomy of algorithmic decision support tasks ($\S$ \ref{sec:tasks_challenges}) and \khedit{to} disentangle sources of target variable bias applicable in each setting} 

\subsection{Scope}\label{sec:scope}

Our framework applies to settings in which a supervised learning model is introduced to augment human decision-making by predicting (i) a future event (e.g., medical \citep{obermeyer2019dissecting}, criminal justice \citep{dieterich2016compas}, child welfare \citep{chouldechova2018case}, or real estate \citep{holstein2023toward, poursabzi2021manipulating} related outcomes); (ii) a subjective human annotation (e.g., perceived content toxicity \citep{gordon2022jury}); or (iii) factual information (e.g., food nutrition \citep{buccinca2021trust}). In these settings, model predictions are combined with human decision-making, either by showing model predictions to a human (i.e., algorithm-in-the-loop \citep{green2019principles}), who makes the final decision, or via a hybrid flow of agency (e.g., deferral-based learning \citep{madras2018predict}, learning with bandit feedback \citep{gao2021human}). Given our focus on prediction-based decision tasks, we do not directly examine decision-support settings involving unsupervised learning (e.g., clustering), tasks relying upon generative models (e.g., text or image generation), or sequential settings with time dependency (e.g., reinforcement learning) in this work.

\subsection{Framework development}\label{sec:process}

Understanding which statistical biases are of concern in a given ADS task requires \textit{examining the historical data generating process} that gave rise to the model training dataset. Causal diagrams, which are graphs that show causal relationships between nodes via connected edges \citep{pearl1995causal}, are tools specifically designed for this purpose. A causal diagram shows each variable being modeled as a \lgdelete{discrete}\textit{node}. Causal connections between nodes are shown via \textit{edges}. If the direction of a causal pathway is known, this is shown via a directed arrow from the parent to child node. An undirected edge is used to connect nodes when the causal direction is unknown or \lgedit{varies across settings described by the diagram} \citep{pearl1995causal}. Our framework \lgedit{introduces} a causal diagram to examine challenges impacting the labels available in data. Therefore, we specifically consider variables (i.e., \textit{nodes}) and relationships (i.e., \textit{edges}) that directly relate to the target variable; we \textit{abstract away} other important factors, such as the training \citep{cai2021onboarding, lai2020chicago, cai2019hello}, decision-making process \citep{green2021algorithmic}, and workflow \citep{green2019principles} of the human decision-makers using the predictive model. While prior work has examined these factors in detail \citep{cai2021onboarding, lai2020chicago, cai2019hello, green2021algorithmic, green2019principles}, our framework foregrounds factors most salient for understanding target variable bias. In $\S$ \ref{subsec:scaffolding}, we outline how our approach can be extended to systematically examine a broader set of components beyond target variables in human-AI decision-making research. 

\begin{figure*}
  \includegraphics[width=.6\linewidth]{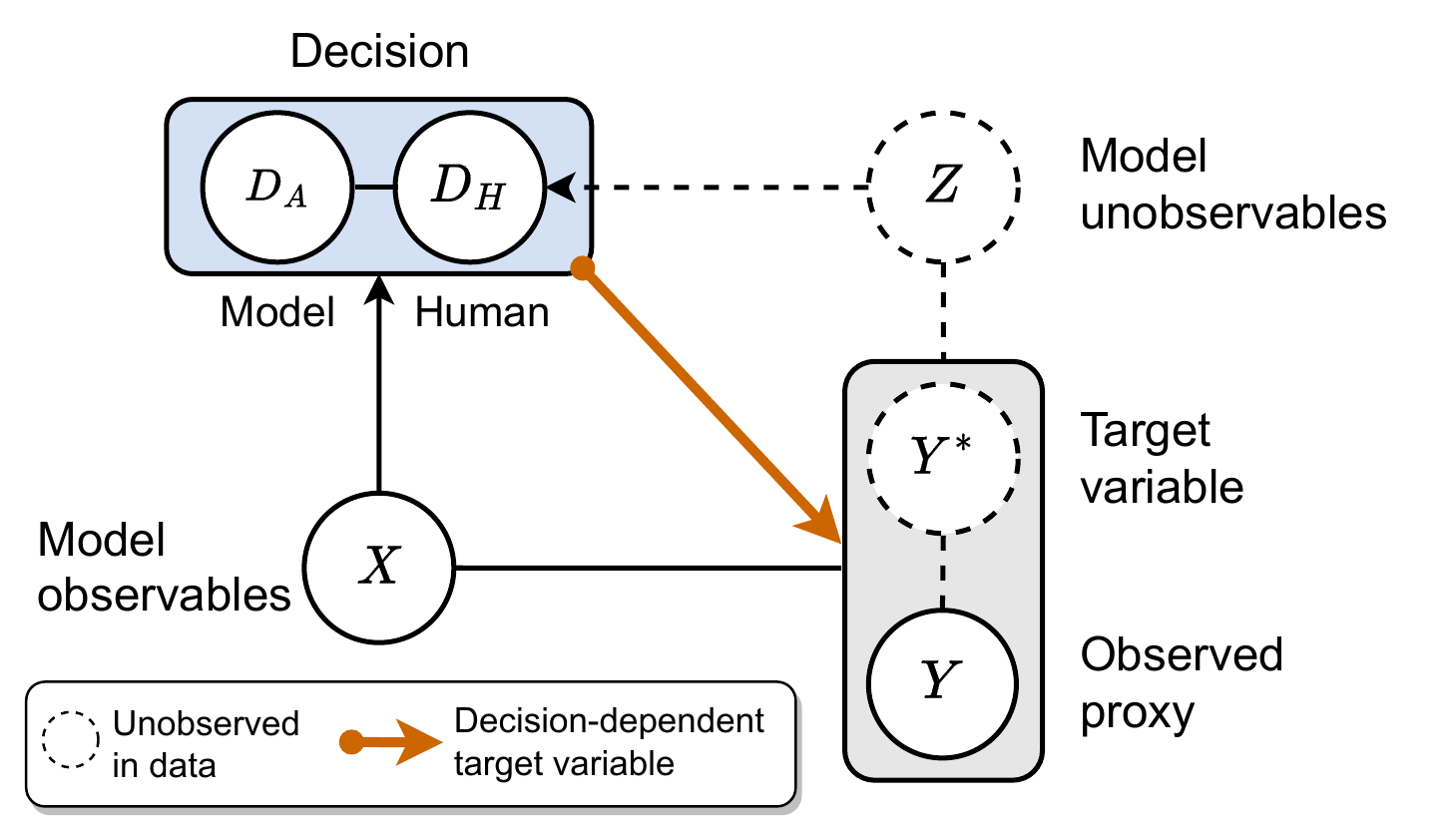}
  \caption{Our causal diagram represents a space of causal graphs, spanning different possible relationships between predictors, decisions, target variables, and their proxies in algorithmic decision support tasks. Edges with directionality that can vary across ADS settings are indicated via undirected edges. Observed variables are shown with solid lines, while unobserved variables are shown in dotted lines. An arrow pointing to a shaded box is shorthand for separate arrows pointing from the source to nodes contained within the box.} 
  \label{fig:dag}
  \Description[A diagram including four nodes arranged in a square structure with directed and undirected lines connecting nodes. An orange line connects the decision and target variable nodes that says ``decision-dependent target variable.'']{A diagram including four nodes arranged in a square structure. Decision node is shown by a blue box in the upper left corner. Inside the blue box are two nodes labeled model and human connected by a dashed undirected line.  Model observables node is shown in the bottom left corner, model unobservables node is shown in the upper right corner. Outcome variables are shown in a grey box in the bottom right corner. Inside the box are two nodes labeled target variable and observed proxy, connected by a dashed undirected line. An orange line points from the decision to target variable nodes labeled decision-dependent target variable. Model unobservables and target variable node have dashed lines, while other nodes contain dotted lines.} 
\end{figure*}

Our causal diagram was developed and refined through an iterative series of discussions among the authors and external researchers spanning a range of disciplines. Based on a review of real-world case studies (see Table \ref{tab:evidence} in Appendix \ref{sec:appendix}), we synthesized candidate causal diagrams that could adequately characterize the target variable of interest across settings, and then stress-tested these diagrams by attempting to identify counterexamples. Through our discussions with external researchers, we also cross-referenced our framework with existing terminology and methods developed in adjacent disciplines, such as medical diagnostic testing, educational assessment, behavioral health, and statistics. 

\subsection{Causal diagram}\label{sec:causal_diagram}
\subsubsection{Diagram structure} Figure \ref{fig:dag} shows our proposed causal diagram, which represents \textbf{a space of directed acyclic graphs (DAGs)} describing the relationship between predictors, decisions, target variables, and their proxies in ADS tasks.

\textbf{Predictors.} $X$ describes \textit{covariates} used to generate model predictions. Covariates are often drawn from administrative data sources (e.g., medical records, lending history) available to an organization for model development. In ADS settings, humans can also make use of \textit{unobserved contextual information} $Z$ while making decisions. For example, a physician might consider real-time medical test results (e.g., electrocardiograms \citep{mullainathan2019machine}) unavailable to a model, while a social worker might weigh contextual factors described via phone calls while deciding whether to recommend investigation of child maltreatment allegations \cite{kawakami2022improving}. In some cases, human decision-makers can also be unaware of a subset of covariates (e.g., due to organizational policy or prohibitively large datasets) \citep{holstein2023toward}. Figure \ref{fig:dag} refers to $X$ and $Z$ as \textit{model observables} and \textit{model unobservables}, respectively, based on whether the predictors are available to a model.

\textbf{Decisions.} The blue shaded box in Figure \ref{fig:dag} shows the joint human-algorithm decision $D$. We decompose this node into separate variables for human decisions $D_H$ and algorithm predictions $D_A$. Prior to deployment of an algorithm, decisions result solely from human judgement ($D_H$). In some cases, \lgedit{post-deployment} decisions \lgdelete{post-deployment} result from humans incorporating predictions into their decision-making (i.e., algorithm-in-the-loop \citep{green2019principles}). In other cases, the joint decisions result from a learned combination of $D_H$ and $D_A$ \citep{gao2021human, madras2018predict, wilder2020learning, tan2018investigating, hilgard2021learning, charusaie2022sample, donahue2022human}. 

\textbf{Target variables.} The node $Y^*$ describes the unobserved target variable of interest to human decision-makers. For example, a model might be introduced to weigh \lgedit{the} risk of unobserved constructs such as ``medical need'' , ``recidivism'', ``creditworthiness'', or ``job performance.'' $Y$ describes the \textit{observed proxy} that is targeted by a model in place of $Y^*$. For example, a model might predict ``cost of medical care'' \citep{obermeyer2019dissecting}, ``re-arrest'' \citep{fogliato2020fairness}, ``loan default'', or ``supervisor performance reviews'' in place of the targets listed previously. The grey box in Figure \ref{fig:dag} represents a \textit{measurement model} mapping the unobserved construct to the observed proxy targeted by a predictive model (see $\S$ \ref{challenge:outcome_measurement_error}).

\textbf{Edges}. We now describe the space of possible relationships \lgedit{connecting nodes} in ADS tasks. Covariates and model unobservables both contribute to human decisions ($D_H$), while algorithmic predictions ($D_A$) are only influenced by covariates ($X$). For example, a physician might make use of medical records ($X$) and real-time test results ($Z$), while an algorithm only has access to medical records ($X$). We show these relationships via directed arrows $X \rightarrow D$ and $Z \rightarrow D_H$. Decisions ($D$) can also influence the target \textit{and} proxy outcomes ($Y^*$, $Y$). \lgedit{For example, enrollment in a medical treatment program can increase medical costs ($Y$) while also improving patient health ($Y^*$).} We show this relationship via the directed arrow $D \rightarrow Y, Y^*$.

The direction of causality between covariates ($X$), unobservables ($Z$), and prediction targets ($Y$ and $Y^*$) can vary across ADS domains. In Figure \ref{fig:dag}, we convey this ambiguity via undirected edges. Causal diagrams for prediction tasks often show covariates ($X$) and unobservables ($Z$) contributing to downstream outcomes ($Y$, $Y^*$) via a domain-specific causal pathway \citep{fairmlook}. In our diagram, this flow of information would be communicated via directed edges from $X$ to ($Y, Y^*$), and from $Z$ to ($Y, Y^*$). However, in some cases, the causal pathway can be \textit{reversed} \citep{hardt2022backward}. For example, this is possible if a patient's unobserved disease status ($Y^*$) contributes to their medical history ($X$) or real-time test results ($Z$). Therefore, bidirectional edges shown in Figure \ref{fig:dag} map to directed edges with directionality that varies depending on the domain. \footnote{In order for the causal diagram to remain valid (i.e., a directed \textit{acyclic} graph), one of the edges connecting nodes must remain disconnected in these settings (i.e., when target variables are decision-dependent and $Y^* \rightarrow X$, $Y \rightarrow X$, $Y^* \rightarrow Z$, or $Y \rightarrow Z$). While this requirement is consistent with the scope of our framework, which considers non-sequential settings, feedback loops are an important factor to consider in sequential settings \cite{ensign2018runaway}.}

\subsection{Mapping algorithmic decision support tasks to sources of target variable bias}\label{sec:tasks_challenges}

\lgdelete{We now leverage our causal diagram to develop a taxonomy of algorithmic decision support tasks (e.g., toxicity detection \citep{carton2020feature}, recidivism prediction \citep{liu2021understanding}, nutrient content prediction \citep{buccinca2020proxy}) studied in prior literature. Our taxonomy contains three distinct \textit{ADS regimes}: tasks with (1) \textit{decision-\khdelete{in}dependent target variables}, (2) \textit{decision-independent target variables}, and (3) \textit{subjective annotations}. Each \khedit{regime maps to a sub-graph}\khdelete{of these regimes map to sub-graphs} of our generalized diagram shown in Figure \ref{fig:dag}. While prior human-AI decision-making studies frequently treat these regimes interchangeably, we show in $\S$~\ref{sec:tasks_challenges} that each setting introduces distinct sources of TVB. Therefore, these regimes should be considered carefully while designing experimental studies and evaluating predictive models.} \lgdelete{The generalized causal diagram we propose in Figure \ref{fig:dag} can be used to articulate key structural differences between decision support tasks studied in human-AI decision-making literature. Identifying the task-specific diagram applicable to a given decision support setting is critical for (1) identifying relevant sources of target variable bias ($\S$ \ref{sec:tasks_challenges}), (2) articulating modeling assumptions ($\S$ \ref{sec:modeling}), and (3) designing ecologically valid experimental studies ($\S$ \ref{sec:experimental_evaluations}).} \lgedit{We now leverage our causal framework to identify sources of target variable bias that can impact predictive models in algorithmic decision support tasks. We begin by introducing two distinct \textit{regimes of ADS tasks} described by our generalized diagram shown in Figure \ref{fig:dag}; those with: (1) decision-dependent target variables, and (2) decision-independent target variables. \textbf{ADS tasks with decision-dependent target variables are subject to more sources of TVB than those with decision-independent target variables}. While it is possible to define \khedit{many} other specific regimes of our generalized diagram (e.g., \khedit{different directions of causality between ($Y, Y^*$) and $X$ or $Z$, or different flows}\khdelete{ a flow} of agency \khedit{between $D_A$ and $D_H$} \khdelete{from $D_A \rightarrow D_H$ versus  $D_H \rightarrow D_A$}), we introduce the distinction between decision-dependent and independent target variables \khedit{here} because it is useful for identifying task-specific sources of TVB.}

\begin{figure*}[t]
  \includegraphics[width=\linewidth]{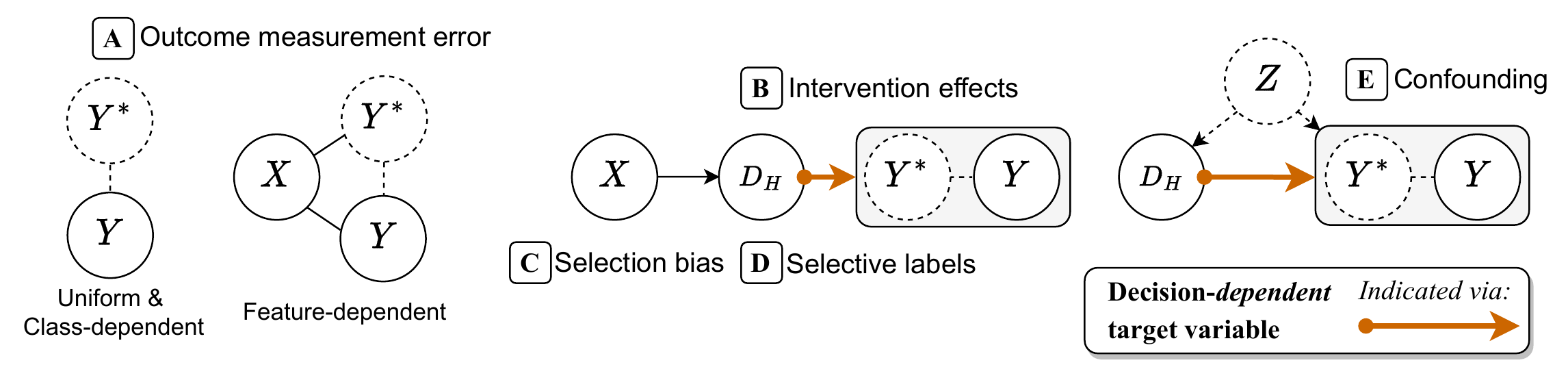}
  \caption{Sub-graphs of the diagram in Figure \ref{fig:dag} introducing statistical biases that impact the target variable $Y^*$. \textit{Outcome measurement error} (\textbf{A}) can occur in settings with both decision-dependent and independent target variables. In decision-dependent settings, \textit{intervention effects} (\textbf{B}), \textit{selection bias} (\textbf{C}), \textit{selective labels} (\textbf{D}), and \textit{confounding} (\textbf{E}) are also of concern. }
  \label{fig:challenges}
  \Description[Four horizontally aligned causal diagrams, each of which contain a subset of nodes contained in the diagram in Figure 1.]{Four horizontally aligned causal diagrams, each of which contain a subset of nodes contained in the diagram in Figure 1. The first two diagrams contain the target and proxy outcomes under the heading outcome measurement error. The center causal diagram contains the labels selection bias, intervention effects, and selective labels surrounding the model observable, human decision, and outcome variables. The right diagram contains the human decision, confounding, and outcome variable nodes. An orange arrow labeled decision-dependent target variable is included in the center and right diagrams.}
\end{figure*}

ADS tasks with \textbf{decision-dependent target variables} occur when \textit{the decision informed by an algorithm also impacts the downstream outcomes} $Y$ and $Y^*$. Real-world ADS deployments often involve prediction tasks with decision-dependent target variables. For example, re-arrest is only observed among defendants released on bail  \citep{kleinberg2018human}, while child welfare screening decisions can influence the risk of adverse care outcomes \citep{coston2020counterfactual}. More generally, \lgdelete{real-world} decisions informed by algorithms often constitute \textit{risk mitigating interventions} (e.g., medical treatments, educational programs) or \textit{opportunities} (e.g., loans, new candidate hires) that change the likelihood of the target outcome (e.g., disease prognosis, educational attainment). Settings with decision-dependent \lgedit{target-variables} include the orange arrow from $D$ to $Y$ and $Y^*$ shown in Figure \ref{fig:dag}. \footnote{\lgedit{ This regime maps directly to the \textit{``predictive optimization''} setting recently studied by \citet{wang2022against} and the discussion of predictive model validity provided by \citet{coston2022validity}.}}

In contrast, the target variable is \textit{not} influenced by the proposed decision in ADS tasks with \textbf{decision-independent target variables}. \khdelete{\khdelete{While t}\khedit{T}his \lgdelete{setting}\lgedit{regime} is uncommon in\lgdelete{real-world} ADS deployments}\khedit{ADS are frequently deployed in the real world with the goal of informing decisions that can change the predicted outcomes. However, lab-based experimental studies of human-AI decision-making}\khdelete{, human-AI decision-making studies} often \lgedit{conduct evaluations via ADS tasks with decision-independent target variables.} For instance, studies have examined models that predict factual content (e.g., food nutrition \citep{buccinca2021trust}) and perceptual information (e.g., counts of objects \citep{park2019slow}, geometric shapes \citep{zhang2022you}). These tasks are decision-independent because the prediction target (i.e., food nutrition, geometric shape) is \textit{not} influenced by the prediction made by a human and/or model. \footnote{\lgedit{ADS tasks in which labels are assigned via human annotations also fall within the decision-independent regime. In these tasks, the ratings of human annotators ($Y$) serve as a proxy for the broader construct of interest ($Y^*$) of interest in the model deployment setting (e.g., comment ``toxicity'' or ``hate speech'' \citep{gordon2022jury} ).}} \lgedit{ADS tasks in the decision-independent regime do not contain the arrow from $D$ to $Y$ and $Y^*$ in Figure \ref{fig:dag}.} 

% \lgdelete{This setting also describes tasks in which the prediction target consists of subjective human annotations (e.g., perceptions of ``toxicity'' or``hate speech''). Here, a specific set of human annotations ($Y$) serve as one possible operationalization of the broader construct of interest ($Y^*$) being targeted by a content moderation model \citep{gordon2022jury, jacobs2021measurement}. Here, annotators serve as a \textit{measurement model} linking the target construct (e.g., toxicity, hate speech) to a proxy (i.e., individual annotation) \citep{jacobs2021measurement}. Because humans can disagree considerably on subjective tasks \citep{gordon2022jury}, labels elicited from specific individuals ($Y$) are an imperfect proxy for the broader construct of interest ($Y^*$). In subjective human annotation tasks, $D$  constitutes decisions made at \textit{runtime} (i.e., after deployment of system predicting the target construct). For example, these decisions can involve removal of content flagged as ``toxic'' or reversal of ``damaging'' Wikipedia edits \citep{halfaker2020ores, kumar2021designing}.}

\lgedit{We now introduce five sources of target variable bias relevant in ADS tasks. \textbf{Outcome measurement error is of concern in \khedit{both}\khdelete{the} decision-dependent \textit{and} decision-independent regimes, while intervention effects, selective labels, selection bias, and confounding bias are only relevant in decision-dependent regime\khedit{s}.} } 

\subsubsection{Outcome measurement error}\label{challenge:outcome_measurement_error}

Human experts often make decisions involving unobserved, latent constructs such as \textit{``recidivism risk''}  and \textit{``job performance.''} These latent constructs are not directly observable in the world, but can be operationalized via a \textit{measurement model} \citep{hand2004measurement, jacobs2021measurement}. Adopting a label observed in data as a proxy for an unobserved latent construct serves as a \textit{de facto} measurement model. For instance, in criminal justice settings, defendant re-arrest is commonly adopted as a proxy for recidivism risk \citep{fogliato2021validity, bao2021s}, while in commercial hiring settings, manager reviews are frequently adopted as a proxy for future job performance.  Outcome measurement error (Figure \ref{fig:challenges}.A) occurs when there is a systematic difference between the target variable of interest to experts and policymakers ($Y^*$) and its operationalization by a proxy ($Y$). This challenge has been extensively documented in judicial \citep{fogliato2020fairness, butcher2022racial}, child welfare \citep{kawakami2022improving, cheng2022disparities}, and hiring \citep{chalfin2016productivity} ADS domains. 

Because proxy labels impacted by measurement error offer an incomplete reflection of the actual goals of human decision-makers, they serve as an incomplete measure of human-AI decision quality. Therefore, before adopting a proxy \lgedit{as a measure of human-AI decision quality}\lgdelete{to evaluate human-AI decision-making}, it is critical to assess whether it serves as a satisfactory approximation of the target variable of interest \lgedit{to humans}. Measurement theory in the quantitative social sciences provides tools to conduct this assessment by weighing the \textit{construct validity} and \textit{reliability} of observed labels \citep{hand2004measurement, jacobs2021measurement} (see $\S$ \ref{subsec:construct_validity}). In practice, measurement error in proxies is often studied via \textit{measurement error models}. These models make \textit{assumptions on the relationship} between the target outcome ($Y^*$) and its proxy ($Y$) (\lgedit{see Appendix \ref{sec:appendix_assumptions}}). \lgedit{Outcome measurement error is of concern in decision-dependent and independent regimes because observed labels can be subject to construct validity and reliability concerns in both settings.}

\subsubsection{Intervention effects}\label{sec:treatment}

In many ADS tasks, decisions serve as \textit{risk mitigating interventions} intended to improve the chances of a favorable policy-relevant outcome \citep{barabas2018interventions, kube2019allocating, coston2020counterfactual}. As a result, past human decisions $D_H$ influence the probability of the target outcome $Y^*$ and its proxy $Y$ (Figure \ref{fig:challenges}.B). However, many existing predictive techniques mistakenly assume that decisions $D$ and outcomes $Y$, $Y^*$ are statistically independent \citep{barabas2018interventions, kube2019allocating, coston2020counterfactual}. This practice can be traced back to formulation of ADS as a prediction-policy problem \citep{kleinberg2015prediction}, in which models are trained to maximize predictive performance with respect to observed outcomes without considering causal effects from $D$ to $Y$ and $Y^*$. Yet, we argue that accounting for the causal connection between decisions and outcomes is of central interest in many ADS tasks. For instance, consider two distinct policy problems that arise in tasks with decision-dependent target variables:

\begin{itemize}
    \item \textbf{Selective Intervention (SI):} In this policy setting, organizations provide resources to individuals who are at \textit{high baseline risk under no intervention}. For example, developers of the Allegheny Family Screening Tool (AFST) introduced the tool with the goal of assessing \textit{``latent risk''} of maltreatment prior to county child welfare interventions \citep{vaithianathan2019allegheny}. Similarly, predictive models have been introduced in educational settings to identify students  at-risk of failing given no tutoring resources \citep{smith2012predictive}. This task requires causal inference because it involves inferring what \textit{would occur} if an individual does not receive the proposed intervention.
    
    \item \textbf{Selective Opportunity (SO):} In this policy setting, an organization grants an opportunity (e.g., a new loan, or pre-trial release on bail) to decision subjects while trying to minimize risk of an adverse outcome (e.g., loan default, recidivism) given an individual receives the opportunity. This prediction task requires causal inference because it involves predicting what would occur under the \textit{hypothetical scenario} that an individual receives the opportunity under consideration.

\end{itemize}

Naively structuring an ADS task as a prediction policy problem in SI and SO settings can lead to misleading assessments of model performance. For example, \citet{coston2020counterfactual} demonstrate that predicting observational modeling and evaluation in SI settings systematically underestimates the risk for high-risk individuals who would respond most favorably to the intervention. The underlying modeling and evaluation challenge introduced by intervention effects stems from the fact that downstream outcomes are only observed under one of the possible decisions \citep{pearl2009causal, perdomo2020performative, johansson2020generalization}. \lgedit{This source of bias is only relevant in the decision-dependent regime because intervention effects are introduced by the connection $D \rightarrow Y, Y^*$.} \lgdelete{This challenge is closely related to an additional source of target variable bias: \textit{selective labels} \citep{lakkaraju2017selective}.}

\subsubsection{Selective labels}\label{sec:selective_labels}

Another challenge introduced by the connection $D \rightarrow Y$, $Y^*$ in \lgedit{the decision-dependent regime} is \textit{selective labels} \lgedit{(Figure \ref{fig:challenges}.D)}. This bias has been widely discussed in connection to pre-trial risk assessments, where recidivism-related proxy outcomes (e.g., re-arrest, failure to appear) are only observed among defendants released on bail \citep{kleinberg2018human, fogliato2021validity, bao2021s, lakkaraju2017selective}. Selective labels also occur in child welfare settings, in which some outcomes (e.g., placement in foster care) are only observed among cases screened-in for investigation \citep{de2018learning}. Selective labels maps directly to \textit{selective intervention} and \textit{selective opportunity} policy problems because we never observe how an individual \textit{would have} benefited from a missed opportunity (SO), or how an intervention \textit{would have} impacted an individual who historically received no additional resources. Selective labels pose the greatest challenge when selection bias was also present in the data generating process. 

\subsubsection{Selection bias}\label{sec:selection_bias}

This bias, which occurs when covariates ($X$) or \lgedit{model} unobservables ($Z$) influenced past decisions ($D$) (Figure \ref{fig:challenges}.C), complicates selective labels and intervention effects. Because a previous decision-making policy may have been more likely to intervene (SI) or grant opportunities (SO) to some sub-populations, these groups may be systematically over- or under- represented in historical outcome data. As a result, ADS models trained on historical data will not perform equally well on all sub-populations during deployment \citep{berk1983introduction}. This effect has been well-documented in recidivism prediction settings, in which models predicting re-arrest outcomes have worse performance among sub-populations historically denied bail \citep{kallus2018residual}. \lgedit{While selection bias can cause challenges in any setting in which data is collected non-randomly \citep{heckman1979sample}, this challenge is compounded in decision-dependent outcome tasks because the connection $X \rightarrow D_H \rightarrow Y^*, Y$ causes selection effects to cascade to selective observation of outcomes $Y$ and $Y^*$}.\lgdelete{While re-weighting techniques can correct for selection bias \citep{swaminathan2015batch}, these approaches typically assume that no model unobservables $Z$ are present, which can be unreasonable in many ADS domains.} The connection between selection bias and other downstream issues (e.g., intervention effects, selective labels) underscores the importance of considering the full data generating process while diagnosing sources of bias impacting proxy labels.

\subsubsection{Confounding bias}\label{sec:confounding} 

This bias occurs when unmeasured variables influence both the treatment and outcome variable \citep{pearl1995causal}. Confounding impacts ADS tasks when unobservables influenced past decisions and downstream outcomes (Figure \ref{fig:challenges}.E) \citep{pearl1995causal}. When confounding impacts ADS models, it is not possible to fully mitigate treatment effects and selective labels via traditional causal inference techniques \citep{pearl2009causal}. Yet, confounding is \textit{not} introduced by model unobservables $Z$ in decision-independent tasks because there is no arrow from decisions $D$ to outcomes $Y$ and $Y^*$. In these tasks, unobservables may serve as an opportunity for complementarity between humans and models arising from asymmetric access to information \citep{holstein2023toward, hemmer2022effect}. Therefore, by mapping an ADS task to its underlying causal diagram and identifying the appropriate task regime, it is possible to identify whether model unobservables pose a treat or opportunity for an ADS deployment.

%% file: Sections/4_modeling.tex
\begin{table*}[t]
\centering

% \resizebox{\columnwidth}{!}{%
{\renewcommand{\arraystretch}{1.2}%
\begin{tabular}{>{\centering\arraybackslash} m{2.3cm}>{\centering\arraybackslash} m{3.4cm}>{\centering\arraybackslash}m{2.8cm}>{\centering\arraybackslash}m{3.1cm}>{\centering\arraybackslash}m{2cm}} 
\toprule

\textbf{Work} & \textbf{Measurement ($F_m$)} & \textbf{Prediction ($F_p$)} & \textbf{Assumptions}  &  \textbf{Bias Mitigated}  \\ 
\hline

\citet{gao2021human}  &  \multirow{5}{1.9cm}[0.1em]{$\hat{Y}^* = F_m[Y]$}  & \multirow{5}{3.0cm}[1em]{\centering \newline\newline$\hat{Y} = \hat{F}_p[X, D_H]$\newline Human decisions $D_H$ available at runtime }  &  \multirow{5}{2.9cm}[0.1em]{\centering Proxy and target variables are equivalent $Y^* = Y$} &   \multirow{5}{2cm}[0.2em]{\centering None}   \\

\citet{madras2018predict}   &   &        &     &     \\ 

\citet{wilder2020learning}    &    &     &   &  \\ 

\citet{tan2018investigating} &   &   &   &  \\ 

\citet{hilgard2021learning}  &   &    &   &    \\ 
\midrule
\citet{de2021leveraging}  & $\hat{Y}^* = F_m[X, D, Y]$, where $\hat{Y}^* = D$ \textit{expert consistency} instances and $Y$ otherwise &  \multirow{5}{3.2cm}[-3.2em]{\centering $\hat{Y} = \hat{F}_p[X]$ \newline Human decisions $D_H$ unavailable at runtime } & Expert consistency assumption & Measurement error, Selection bias \\ 
% \cmidrule{1-2}\cmidrule{4-5}

\citet{lakkaraju2017selective} & $\hat{Y}^* = F_m[Y]$ &   & Heterogeneous acceptance rates & Selection bias \\ 

% \cmidrule{1-2}\cmidrule{4-5}
\citet{coston2020counterfactual}  & $\hat{Y}^* = F_m[Y_d]$, where $Y_d$ is a potential outcome  &  &  Causal identifiability conditions & Intervention effects \\ 

% \cmidrule{1-2}\cmidrule{4-5}
\citet{coston2020confounding}                                              & $\hat{Y}^* = F_m[Y_d, Z_r]$, where $Y_d$ is a potential outcome  &  &  Causal identifiability conditions & Intervention effects, Confounding   \\ 

% \cmidrule{1-2}\cmidrule{4-5}
\citet{wang2021fair}  & $\hat{Y}^* = F_m[Y]$,  where $Y$ error is group-dependent  &  &  Confident learning assumptions (see \citep{northcutt2021confident}) & Measurement error \\ 

\midrule
\midrule

\textbf{Label noise} \citet{menon2015learning}  & $\hat{Y}^* = F_m[Y]$, where $Y$ class-conditional or positive and unlabeled & ERM with surrogate loss (see \citep{natarajan2013learning}) & Weak separability  & Measurement error \\

\midrule
\textbf{Latent Class Analysis} \citet{mccutcheon1987latent}  & $\hat{Y}^* = F_m[Y]$, where $Y=\{Y^1, ..., Y^K\}$ are independent factors & 3-step LCA with covariates (see \citep{vermunt2010latent}) & $Y^i \bigCI Y^j \mid Y^*$  & Measurement error \\

\midrule
\textbf{Hui-Walter Framework} \citet{hui1980estimating}  & $\hat{Y}^* = F_m[Y]$, where $Y=\{Y^1, ..., Y^K\}$ are diagnostic tests & N/A & Test Se/Sp identifiability assumptions & Measurement error\\ 

\bottomrule
\end{tabular}
\caption{Taxonomy of measurement and prediction approaches. Top: methods proposed in ADS literature. Bottom: 
 methods applied in machine learning, social sciences, and bio-statistics.}\label{tab:modeling_table}
}

\end{table*}

\section{Model Development}\label{sec:modeling}

We now provide a framework for specifying target variable assumptions during predictive model development. We argue that predictive modeling for ADS involves two distinct steps: \textit{measurement} and \textit{prediction}. During the measurement step, tool developers construct a \textit{measurement model} that operationalizes the target \lgdelete{outcome}\lgedit{variable} of interest $Y^*$ using readily available datasets. During the second step, tool designers train a \textit{prediction model} that targets the \textit{proxy outcome} returned by the measurement model. We now discuss each of these modeling steps in detail.

\subsection{Measurement model}\label{sec:measurement}

During the measurement step, the unobserved outcome of interest ($Y^*$) is approximated using historical data from the causal diagram in Figure \ref{fig:dag}. This step involves establishing a \textit{measurement hypothesis} ($\hat{Y}^*$) using observed information: covariates $X$, past decisions $D$, and one or more outcome proxies $Y$. In some settings, a subset of unobservables are available during model development, but unavailable in during deployment. Such runtime confounders $Z_r \subseteq Z$ can occur when protected attributes (e.g., race, gender) are available during development, but not in deployment for legal reasons \citep{diana2022multiaccurate, coston2020runtime}. Given information $X$, $Z_r$, $D$, $Y$ recorded in existing data, we can construct a measurement model approximating \lgedit{the target variable} $Y^*$: 
\begin{equation}\label{eq:measurement}
  \hat{Y}^* = F_m[X, Z_r, D, Y]
\end{equation}

Unlike statistical models commonly used in machine learning contexts, a measurement model cannot be learned from past data because the target outcome $Y^*$ is unobserved. Instead, $F_m$ relies on \textit{measurement assumptions} concerning the relationship between the unobserved outcome of interest and recorded information available for modeling. Therefore, it is not possible to assess the quality of $\hat{Y}^*$ by comparing against held-out data, as is common in prediction settings. Instead, evaluating measurement models requires a multifaceted approach, including assessments of construct validity, synthetic experiments, sensitivity analyses, and other evaluation strategies described in $\S$ \ref{sec:measurement_model_evaluation}.  

All predictive models in ADS introduce a measurement model. However, this model is often \textit{implicitly defined} and makes tacit assumptions on the relationship between available data sources ( $X$, $Z_r$, $D$, $Y$) and the target variable ($Y^*$). Table \ref{tab:modeling_table} provides a detailed list of the measurement models assumed by existing ADS approaches. This table reifies often-implicit measurement assumptions adopted by prior work. In the bottom three rows, we apply our taxonomy to workhorse methods used in machine learning \citep{menon2015learning, natarajan2013learning}, quantitative social sciences \citep{mccutcheon1987latent}, and bio-statistics \citep{hui1980estimating} literature. \lgedit{The \textit{Bias Mitigated} column of Table \ref{tab:modeling_table} refers to the source of TVB addressed by the modeling technique. For instance, we mark ``None'' for \citet{wilder2020learning} and \citet{madras2018predict} because these approaches are not designed to mitigate any TVB sources listed in $\S$ \ref{sec:tasks_challenges}.}

\subsection{Prediction model}\label{sec:prediction}

After establishing a measurement model to estimate $Y^*$ given ($X$, $Z_r$, $D$, $Y$), tool designers then train a \textit{prediction model} for use in decision-support settings. This prediction model takes observed covariates ($X$) and predicts the measurement hypothesis ($\hat{Y^*}$) established during the preceding measurement step. Because $Z_r$ and $Y$ are unavailable during deployment, these are not included in the prediction model. Most often, prediction models do not assume human decisions $D$ are available at runtime (i.e., algorithm-in-the-loop \citep{green2019principles}). However, in some more nuanced decision-making workflows, models may also assume that human decisions are available at run-time as an additional input (i.e., \citep{madras2018predict, wilder2020learning, gao2021human, tan2018investigating}). Given $X$ and optionally $D$ available at runtime, the prediction model estimates the \textit{measurement hypothesis} $\hat{Y}^*$:
\begin{equation}\label{eq:prediction}
  \hat{Y} = \hat{F}_p[X, D]
\end{equation}

Whereas a measurement model is \textit{constructed via measurement assumptions}, the prediction model $\hat{F}_p$ is a \textit{learned mapping} from $X$ (and in some cases $D$) to the measurement hypothesis $\hat{Y}^*$. Therefore, it is appropriate to evaluate generalization of $\hat{F}_p$ to held-out data via the standard slate of evaluation metrics (e.g., accuracy, AU-ROC, or statistical fairness measures). Critically, this evaluation is conducted with respect to the measurement hypothesis established during the measurement step ($\hat{Y}^*$) rather than the target outcome ($Y^*$) directly. Thus, showing strong performance of $\hat{F}_p$ \textit{is not sufficient to claim a model generates valid predictions for the target outcome} $Y^*$. 

\subsection{Measurement model evaluation}\label{sec:measurement_model_evaluation}

\lgdelete{Because the target outcome $Y^*$ is unobserved, m}\lgedit{M}easurement model evaluation requires a holistic, multifaceted approach leveraging converging sources of evidence. Informed by methods used in statistics, quantitative social sciences, and learning sciences, we provide a recommended set of approaches for validating measurement models in ADS tasks. 

\subsubsection{Construct reliability and validity}\label{subsec:construct_validity}

Measurement theory offers a comprehensive set of criteria for assessing the quality of a measurement model. \textit{Construct reliability} describes the degree to which a latent phenomena is consistently reflected by a measurement model (e.q. \ref{eq:measurement}) over time. Threats to construct reliability have been well documented in settings in which target variables are assigned via subjective human annotations. In these settings, assignment of target outcomes can vary substantially based on rater identity \citep{denton2021whose, denton2021whose}, context \citep{pavlopoulos2020toxicity}, and specification of the annotation protocol \citep{poletto2021resources}. Construct validity describes the extent to which a measurement model adequately captures an unobserved phenomenon of interest. Thus, while construct reliability is roughly analogous to the notion of statistical variance in $F_m$, construct validity is analogous to statistical \textit{bias} in $F_m$ \citep{jacobs2021measurement}. We refer the reader to \citep{coston2022validity} for a detailed discussion of sub-components of construct reliability and  validity that pertain to risk assessment development and evaluation. 

\subsubsection{Outcome cross-validation}
In many ADS domains, multiple proxies are available that are believed to be related to the target outcome of interest. In the criminal justice domain, courts often track multiple recidivism-related outcomes (e.g., 2-year general and violent recidivism, failure to appear). In the child welfare domain, government agencies may track substantiation of abuse allegations, acceptance for welfare services, agency re-referral, placement in foster care, and hospitalization \citep{vaithianathan2019allegheny}.  When multiple reference outcomes are available, \textit{outcome cross-validation} can be used to train a model to predict one proxy, then evaluate this model on a slate of \textit{additional reference variables} that \lgedit{domain experts}\lgdelete{modelers} expect may be reasonable proxies for the outcome of interest. If targeting a proxy \textit{also} results in strong performance across other reference variables, this provides evidence suggesting that a proxy may serve as a suitable measurement model. Outcome cross-validation has been independently used by analyses of proxy outcomes in learning analytics \citep{rachatasumrit2021toward}, criminal justice \citep{kleinberg2018human}, child welfare \citep{de2021leveraging}, and healthcare \citep{obermeyer2019dissecting}. Special cases of outcome cross-validation map to sub-components of construct validity. For example, a model demonstrates \textit{predictive validity} if its predictions correlate with a reference outcome known to be related to the construct of interest \citep{hand2004measurement}. A model demonstrates \textit{discriminant validity} if its predictions are \textit{not} correlated with a conceptually distinct outcome.

\begin{figure*}
\begin{minipage}{0.44\textwidth}
\begin{figure}[H]
    \centering
    \includegraphics[scale=0.6]{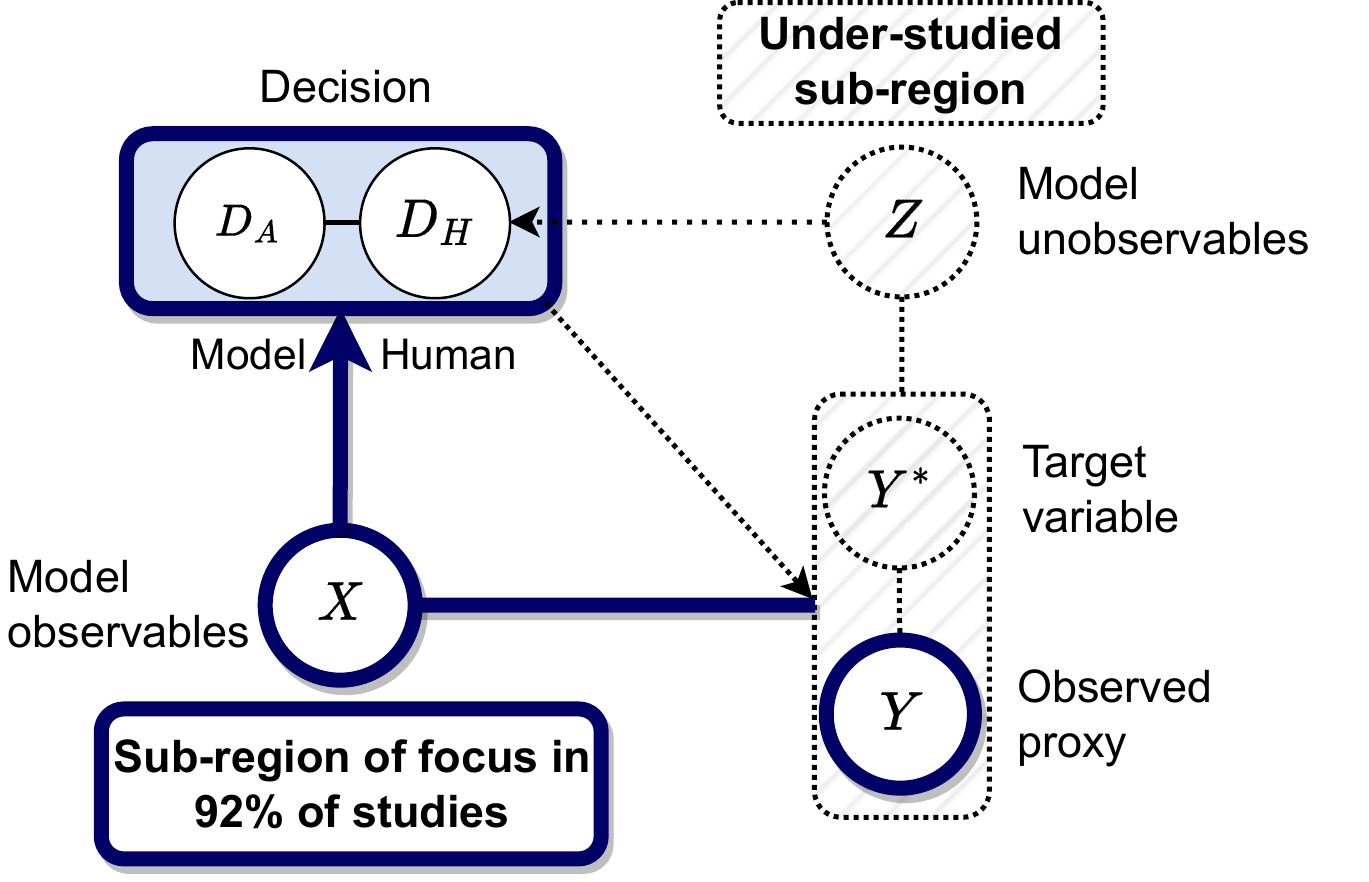}
    \caption{66 of the 72 studies ($\approx92\%$) in our review examine a narrow sub-region of our proposed causal diagram.} 
    \label{fig:experimental_studies_dag}
      \Description[The causal diagram from Figure 1 is shown with the blue shaded decision region, model observables node, and observed proxy node bolded in a dark blue border. Region is labeled ``sub-region of focus in 92 percent of studies''.]{}
\end{figure}
\end{minipage}
\hfill
\begin{minipage}{0.48\textwidth}
\begin{table}[H]
    \centering
    \resizebox{\textwidth}{!}{\begin{tabular}{>{\centering\arraybackslash} m{3cm}>{\centering\arraybackslash} m{2cm}>{\centering\arraybackslash}m{2.5cm}} 
        \toprule
        
        \textbf{Work} & \textbf{Setting} &  \textbf{Sub-region of causal diagram}  \\ 
        \midrule
        
        \citet{hemmer2022effect} & \multirow{2}{2cm}[-.2em]{\centering House price prediction} & \multirow{2}{2.5cm}[-.1em]{\centering Unobservables:\\  $D \leftarrow Z \rightarrow  Y$} \\
        
        \citet{holstein2023toward} & & \\
        \midrule
        \citet{gordon2022jury} & Toxicity detection & Measurement error: $Y^* \rightarrow Y$  \\
        \midrule
        \citet{peng2019you} & Hiring & Selection bias: $X \rightarrow D$  \\
        \midrule
        \citet{green2021algorithmic} & \multirow{2}{2cm}[-.2em]{\centering Judicial} & \multirow{2}{2.5cm}[-.1em]{\centering Omitted payoffs: $Q \rightarrow D_H$}  \\
        \citet{fogliato2021impact} &  & \\
        \midrule
        \end{tabular}}
    \caption{Experimental studies examining the under-studied sub-region provided in Figure \ref{fig:dag}.}\label{tab:tvb_studies}
\end{table}

\end{minipage}

\end{figure*}

\subsubsection{Sensitivity analyses}

Sensitivity analyses enable assessing the degree of \textit{measurement model misspecification} permissible before evaluation of a \textit{prediction model} is invalidated. This technique has traditionally been applied in causal inference settings to estimate the magnitude of unobserved confounding necessary to invalidate a treatment effect estimate \citep{rosenbaum2005sensitivity, diaz2013sensitivity}. More recently, sensitivity analyses have been developed for predictive model evaluation. For instance, \citet{fogliato2020fairness} proposed a sensitivity analyses framework that examines the degree of outcome measurement error permissible before fairness-related analyses are invalidated. Future work in ADS would benefit from sensitivity analysis frameworks that examine multiple sources of target variable bias in parallel. 

\subsubsection{Synthetic evaluation}

A limitation of leveraging real-world datasets for measurement model validation is that one never knows the actual relationship between $Y$ and $Y^*$ in naturalistic data. Model-level evaluations in ADS typically circumvent this issue via \textit{synthetic evaluations} \lgedit{which} test whether proposed approaches are robust to experimentally manipulated bias \citep{de2021leveraging, coston2020counterfactual, menon2015learning, wang2021fair}. Yet, synthetic evaluations require assuming a specific measurement error model. If the data generating process adopted by a synthetic evaluation does not reflect real-world conditions, this can lead to \lgedit{overconfidence in model performance in more realisitic settings}. This concern is salient because synthetic evaluations are often designed with bespoke data generating processes intended to highlight the specific challenge being addressed by the technique. 

\subsubsection{The Oracle Test}

\citet{chouldechova2018case} propose a conceptual tool called the \textit{``Oracle Test''}, which can surface unforeseen sources of target variable bias. This thought experiment supposes that we have access to an oracle model that can predict a proxy with perfect accuracy. The key question posed by this test is: \textit{``What concerns remain given access to such an oracle?''}. Because we have a ``perfect'' prediction model (e.q. \ref{eq:prediction}), remaining concerns are often related to measurement and validity (e.q. \ref{eq:measurement}). For example, \citet{chouldechova2018case} surface concerns related to measurement error when they apply the Oracle Test to examine RAIs designed for ADS models deployed in the child welfare domain. \citet{green2021algorithmic} also leverage the Oracle Test by arguing that improvements to predictive accuracy do not equate to improved public policy outcomes when competing factors in addition to risk (i.e., defendant liberty) \lgedit{are overlooked}. \lgdelete{play a role in judicial payoffs.}

%% file: Sections/5_experiments.tex
\section{Assessing Gaps and Opportunities for Experimental Research }\label{sec:experimental_evaluations}

\lgedit{In this section, we leverage our framework to assess the extent to which existing lab-based studies consider sources of target variable bias ($\S$ \ref{subsec:mapping_existing}). Our analysis finds systematic gaps in our current understanding of human-AI decision-making in light of TVB\khedit{.}\khdelete{, which may impact the ecological validity of prior lab-based studies.} \khedit{We then}\khdelete{Therefore, we} show how our framework can be used by researchers to assess threats to the \khedit{ecological validity and} generalizability of lab-based studies ($\S$ \ref{subsec:ecological}). We conclude by discussing opportunities to use our methodology to explore a broader space of open challenges in human-AI decision-making research ($\S$ \ref{subsec:scaffolding}).} \lgdelete{Our \lgdelete{causal} framework can be used to assess the extent to which existing \lgdelete{in vitro} \lgedit{lab-based} studies consider measurement error, intervention effects, and related challenges ($\S$ \ref{subsec:mapping_existing}). By mapping existing studies to corresponding sub-regions of our proposed diagram, we surface \lgedit{key gaps} in our current understanding of human-AI decision-making. Our causal diagram can be used to evaluate whether experimental findings are likely to hold \textit{ecological validity} in a given real-world ADS deployment ($\S$ \ref{subsec:ecological}). More broadly, the methodology of causal diagrams can also serve as a scaffolding for structuring empirical knowledge beyond concerns related to target variable bias ($\S$ \ref{subsec:scaffolding})} \lgedit{In Appendix \ref{subsec:appendix_experimental_coding}, we provide a resource that helps researchers apply our causal framework to examine the design and ecological validity of several experimental human-AI decision-making studies.}

\subsection{Mapping existing experimental study designs to our causal diagram}\label{subsec:mapping_existing}

To assess the extent to which existing studies examine factors related to target variable bias in their study design, we revisit a comprehensive literature review conducted by \citet{lai2021towards} through the lens of our causal diagram (Figure \ref{fig:dag}). \citet{lai2021towards} review over one hundred experimental studies of algorithm-assisted decision-making published in premiere venues between 2018 and 2021. Our follow-up analysis extends this review to studies published in 2022 at the same set of venues, in addition to \lgdelete{other recently published}\lgedit{recent} pre-prints. We further limit selection criteria applied by \citet{lai2021towards} to studies examining prediction-based decision-making settings (i.e., scope  outlined in $\S$ \ref{sec:scope}). Thus, we exclude studies included in the initial review with a focus on NLP-related tasks.

\textbf{Our analysis finds that 66 out of 72 ($\approx$ 92\%) studies satisfying our criteria conduct experimental evaluations focusing on a narrow sub-graph of our causal diagram}. These studies investigate a modification to the joint decision-making process (i.e., the blue $D_H$ and $D_A$ region) using observed attributes $X$ and an outcome proxy $Y$ (Figure \ref{fig:experimental_studies_dag}). Such studies assume that (1) the target variable and proxy are equivalent (i.e., no measurement error \lgdelete{or validity concerns}), (2) all predictors are observed by both the algorithm and the human (i.e., no model unobservables), and (3) decisions and outcomes are unrelated (i.e., no intervention effects).

Six of the remaining studies we review examine different sub-regions of the causal diagram described in Figure \ref{fig:dag}. Table \ref{tab:tvb_studies} groups these studies by the sub-region under study, including unobservables \citep{hemmer2022effect, holstein2023toward}, measurement error \citep{gordon2022jury}, selection bias \citep{peng2019you}, and omitted payoffs \citep{green2021algorithmic, fogliato2021impact}. While these studies offer early insight into how target variable bias can impact algorithm-assisted human decision-making, our empirical understanding of these challenges remains limited compared to the joint human-AI decision region investigated by $\approx92\%$ of studies. Critically, no work in our review experimentally manipulated factors related to \textit{intervention effects} or examined \textit{multiple intersecting sources of bias} in parallel. Given the prevalence of compounding challenges in real-world settings, \textbf{this gap opens a broad space of open questions and future opportunities for human-AI decision-making research}. 

\vspace{-1mm}
\subsection{Assessing the ecological validity of lab-based studies}\label{subsec:ecological}

The gap we identify between \lgedit{real-world} challenges and \lgedit{lab-based} \lgdelete{in vitro} studies (i.e., Figure \ref{fig:experimental_studies_dag}) carries implications for the ecological validity of experimental studies. Threats to ecological validity may be most acute when findings from a controlled study conducted under simplified conditions are \textit{generalized} to real-world ADS deployments in which multiple sources of target variable bias are present. In these settings, measurement error and intervention effects could impact whether findings gathered via controlled experiments also apply in more complex real-world conditions. 

Fortunately, our causal diagram provides a tool for assessing whether findings from a \lgedit{lab based} study are likely to generalize to a given \lgedit{real-world} \lgdelete{\textit{in vivo}} ADS tool deployment. The first step in this process involves mapping the ADS task to its corresponding \lgdelete{DAG} \lgedit{regime identified in $\S$ \ref{sec:tasks_challenges}} \lgdelete{(e.g., identifying whether the task involves decision-dependent vs. decision-independent outcomes)}. Next, based on domain expertise, one can identify whether different sources of bias are likely to be relevant in the given real-world deployment. For instance, a model deployed to allocate tutoring resources (i.e., a decision-dependent task) may need to account for measurement error in learning outcomes and intervention effects from historical tutoring decisions. In contrast, a model deployed for a perceptual assessment task (e.g., predicting current forest cover from satellite \lgedit{imagery \citep{wang2021explanations}; a task with decision-independent outcomes}) may not need to address these concerns. After identifying the appropriate \lgedit{ADS regime} and relevant sources of bias, one can assess whether an experimental study is likely to generalize to this setting by examining whether the study used a similar prediction task (e.g., also tested decision-dependent or decision-independent outcomes).  

To demonstrate how causal diagrams can be used to assess ecological validity of lab-based studies, consider a previous lab-based assessment conducted by \citet{park2019slow}. This study -- which is sampled from the 66 studies covered by the blue sub-region of the causal diagram provided in Figure \ref{fig:experimental_studies_dag} -- examines whether introducing a delay between when humans view observed features $X$ and algorithmic recommendations $D_A$ improves their performance on a perceptual jellybean counting task. Because the true quantity of jellybeans does not depend on the decision under consideration, this \lgedit{study involves a task from the decision-independent outcome regime} \lgdelete{setting involves decision-independent outcomes}. Further, the influence of human-only observed attributes $Z$ and measurement error is limited in this task. Therefore, findings from this work may most readily generalize to real-world decision-making settings with limited interference from measurement error, model unobservables, and intervention effects.

\subsection{Scaffolding a science of human-AI decision-making}\label{subsec:scaffolding}

Our work leverages causal diagrams to characterize sources of bias impacting target variables. However, beyond this focus, causal diagrams also offer a powerful scaffolding for studying other aspects of human-AI decision-making\lgedit{, such as the joint human-AI decision-making process (i.e., the $D$ node in our framework)}. \lgdelete{While our analysis is \lgedit{focused on target variables} we model the target variable ($Y^*$) endogenously as a function of $(X, Z, D, Y)$, one could also construct a causal diagram examining the joint human-AI decision-making process.} For example, \citet{green2021algorithmic} specify a \lgedit{causal diagram} that models how \lgdelete{humans}\lgedit{judges} weigh risk against other competing factors (e.g., culpability, value of defendant freedom) during pre-trial release decisions. The authors then experimentally verify a \textit{hypothesized} edge in this \lgedit{causal diagram} via a controlled online study. Through a series of such studies, it may be possible to develop a more generalized \textit{theory} of AI-assisted human decision-making across decision support tasks. This process of specifying, testing, and refining causal models is central to existing empirical disciplines, including psychology and sociology \citep{pearl1995causal}. 

%% file: Sections/6_discussion.tex
\section{Discussion}

Our work surfaces a disconnect between the challenges that arise in \textit{real-world deployments} of algorithmic systems versus \textit{current research practices} (i.e., experimental study designs, modeling assumptions, measures of human-AI decision quality) adopted in the human-AI decision-making \lgedit{literature}. \lgedit{Left unaddressed, current gaps in this literature can amount to substantive downstream harms. For instance, while prior studies of real-world ADS tool deployments have surfaced patterns of \textit{apparent} human under-reliance arising from imperfect prediction targets \citep{kawakami2022improving, cheng2022disparities, stevenson2022algorithmic}, no experimental human subjects studies to date have examined how to disentangle \lgedit{warranted} skepticism in a misaligned model versus \textit{unwarranted} under-reliance due to algorithm aversion. Absent such knowledge, organizations may continue to pressure domain experts to rely upon flawed predictive models \citep{kawakami2022improving}, which have been shown to misallocate of medical resources \citep{obermeyer2019dissecting} and perpetuate historical patterns of bias \citep{fogliato2021validity, akpinar2021effect, bao2021s} (see Table \ref{tab:evidence} in Appendix \ref{sec:appendix} for additional examples of real-world harms introduced by TVB).}

Our work provides a critical first step for addressing this disconnect by clarifying the relationship between measurement error, intervention effects, unobserved confounding, \lgedit{selective labels, and} selection bias via intuitive causal diagrams. Going forward, we hope that this framework will support more comprehensive assessment of modeling techniques ($\S$ \ref{sec:modeling}) and empirical human subjects studies ($\S$ \ref{sec:experimental_evaluations}) designed to facilitate human-AI decision-making. \lgdelete{that investigate the implications of bias in algorithm-assisted \textit{human} decision-making.} \lgedit{However, further work is needed to gain a comprehensive understanding of the sources and implications of target variable bias in human-AI decision-making research.} \lgdelete{our work is only the first step towards comprehensively addressing target variable bias through human-AI decision-making research.}

In particular, future research should develop holistic measures of decision-quality that reflect factors beyond statistical performance computed via a single outcome proxy. These measures should reflect both \textit{process-oriented} considerations (i.e., how multiple decision-relevant factors are weighted \citep{green2019principles}, and adherence to procedural, interpersonal, and informational justice) in addition to \textit{outcome-oriented} considerations (i.e., whether a decision led to a beneficial outcome). Where possible, outcome-related measures should draw upon \textit{multiple decision-relevant proxies} to better account for limitations of adopting any single proxy in isolation. While this practice is standard in disciplines such as learning sciences, diagnostic medical testing, and psychology, to date, human-AI decision-making research has primarily adopted outcome-oriented measures that hinge upon on a single potentially flawed proxy. 

Our work also motivates exciting new lines of human-AI decision-making research. \lgedit{For instance, our review of prior modeling approaches finds that, while many techniques have been designed to address a subset of model reliability challenges (Table \ref{tab:modeling_table}), few examine how various sources of target variable bias compound in real-world deployment scenarios.} Additionally, our review of experimental human subjects research provides a set of tools for (i) identifying open empirical questions (i.e., Figure \ref{fig:experimental_studies_dag}), (ii) designing studies with robust ecological validity, and (iii) synthesizing findings from multiple experimental studies into a complete scientific understanding of human-AI decision-making. We hope that our work will raise awareness of target variable bias in the human-AI decision-making research community and spur efforts to better align research practices with the complex challenges encountered in real-world ADS deployments.

\begin{acks}

We thank Stevie Chancellor, Steven Dang, Maria De-Arteaga, Shamya Karumbaiah, Ken Koedinger, and \lgedit{annonymous reviewers} \lgdelete{attendees of the NeurIPS 2022 Workshop on Human-Centered Machine-Learning} for their helpful feedback. We acknowledge support from the UL Research Institutes through the Center for Advancing Safety of Machine Intelligence (CASMI) at Northwestern University, the Carnegie Mellon University Block Center for Technology and Society (Award No. 53680.1.5007718), and the National Science Foundation Graduate Research Fellowship Program (Award No. DGE-1745016). \lgedit{ZSW is supported in part by the NSF FAI (Award No. 1939606), a Google Faculty Research Award, a J.P. Morgan Faculty Award, a Facebook Research Award, an Okawa Foundation Research Grant, and a Mozilla Research Grant.}

\end{acks}

%% file: Sections/7_appendix.tex
\appendix
\section{Appendix}\label{sec:appendix}

\begin{table*}[th]
\centering

{\renewcommand{\arraystretch}{1.1}%
\begin{tabular}{>{\centering\arraybackslash} m{2.8cm}>{\centering\arraybackslash} m{1.2cm}>{\centering\arraybackslash}m{10cm}} 
\toprule

\textbf{Work} & \textbf{Domain}  & {\centering\textbf{Bias Reported}}  \\ 
\hline

 \citet{kleinberg2018human}  &   \multirow{4}{1.2cm}[-.6em]{\centering Judicial}  & {\raggedright Unobservables, selection bias, and outcome measurement error impacting pre-trial risk assessments }\\

% \citet{fogliato2021validity}   &          &  {\raggedleft Measurement error introduced by adopting re-arrest as a proxy for re-offense}  \\ 

\citet{bao2021s}      &     &  {\raggedright Selection bias and measurement error impacting by recidivism RAIs } \\ 

\citet{butcher2022racial} &     & {\raggedright Measurement error in re-arrest proxy outcomes introduced by differential arrest rates among Black and white defendants } \\ 

\midrule

\citet{kawakami2022improving} \citet{cheng2022disparities}  &  Child Welfare  &  

\multirow{1}{10cm}[.9em]{\raggedright Documents social worker concerns that measurement error and unobservables impact the quality of ADS predictions}
  \\ 
\midrule

\citet{obermeyer2019dissecting}  &  \multirow{3}{1.3cm}[-1.1em]{\centering Medical}  &  {\raggedright Measurement error arising from adopting \textit{``cost of care''} as a health proxy } \\

 \citet{mullainathan2017does}  &   &  {\raggedright Measurement error introduced when using medical records as a proxy for stroke outcomes} \\

 \citet{mullainathan2019machine}  &   & {\raggedright  Unobservables, selection bias, and measurement error in clinical decision support } \\

\midrule

 \citet{chalfin2016productivity}  & Hiring & {\raggedright  Omitted payoffs, measurement error, and selection bias arising in \textit{teacher value-add} proxy used for educator hiring } \\ 
 
\bottomrule
\end{tabular}
}
\caption{Documented examples of target variable bias impacting predictive models across numerous ADS domains.}
\label{tab:evidence}
\end{table*}

\subsection{Descriptions of widely-studied outcome measurement error models}\label{sec:appendix_assumptions}

\begin{itemize}
    \item \textbf{Uniform} error assumes that the target outcome is randomly corrupted by additive noise (i.e., $Y^* = Y + \epsilon$) \citep{pischke2007lecture}. This setting is also sometimes called \textit{classical measurement error} in statistics and economics. Because it is possible to learn an unbiased estimate for $Y^*$ given proxy labels $Y$ in uniform error settings \cite{menon2015learning}, this error model poses fewer threats to validity than others discussed below.
    \item \textbf{Class-dependent} error assumes that positive and negative target outcomes are misclassified at different rates. As with uniform error, measurement error in this setting is uncorrelated with co-variates ($Y \CI Y^* | X$) and model unobservables ($Y \CI Y^* | Z$). This model is referred to as \textit{asymmetric} or \textit{class conditional label noise} in machine learning literature \citep{scott2013classification}, and \textit{nondifferential mismeasurement} in statistics and epidemiology \citep{ogburn2013bias}. In contrast to uniform error settings, training a model to predict a proxy ($Y$) impacted by class dependent error \textit{will} lead to biased estimates for the target outcome ($Y^*$) when optimizing accuracy \cite{menon2015learning}.
    \item \textbf{Feature-dependent} error occurs \textit{differentially} across sub-populations based on co-variates ($Y \nCI Y^* | X$) or model unobservables ($Y \nCI Y^* | Z$). This model is called \textit{differential mismeasurement} in statistics and \textit{feature-dependent label noise} in machine learning literature \citep{frenay2013classification}. This setting is also called \textit{group-dependent error} when the covariate in question is a protected attribute (e.g., gender, race) \citep{wang2021fair}. Group-dependent error inherits modeling challenges arising in the class dependent case, and has been tied to disparities in criminal justice \citep{obermeyer2019dissecting} and medical \citep{akpinar2021effect} outcomes in real-world deployments of ADS tools.    
\end{itemize}

% EXTENDED VERSION
Human-AI decision-making research also stands to benefit from existing \textit{methodologies} designed to characterize measurement error in other disciplines. Latent Class Analysis (LCA) is an approach used in psychology and political science to identify latent sub-populations in data that are believed to carry an unobserved characteristic (e.g., personality, political ideology, or disease status) \citep{weller2020latent}. LCA estimates a set of conditional probabilities mapping multiple discrete \textit{factors} (i.e., \textit{proxies}) to a binary latent variable (e.g., \textit{target outcome}). While LCA is tailored to discrete latent variables, other structural equation models (i.e., factor analysis \cite{harman1976modern}) are designed for continuous latent variables. Within biostatistics, the Hui-Walter framework is used to estimate the sensitivity and specificity of diagnostic tests in the absence of a gold standard \citep{hui1980estimating}. Given multiple proxies, Hui-Walter can therefore be adapted to estimate the sensitivity and specificity of each proxy. Like all measurement models, LCA and Hui-Walter make assumptions on the relationship between the target outcome and its proxy. Table \ref{tab:modeling_table} states these assumptions in the context of our measurement model taxonomy.

\clearpage

\begin{figure*}
\centering
 \makebox[\textwidth]{\includegraphics[width=\textwidth]{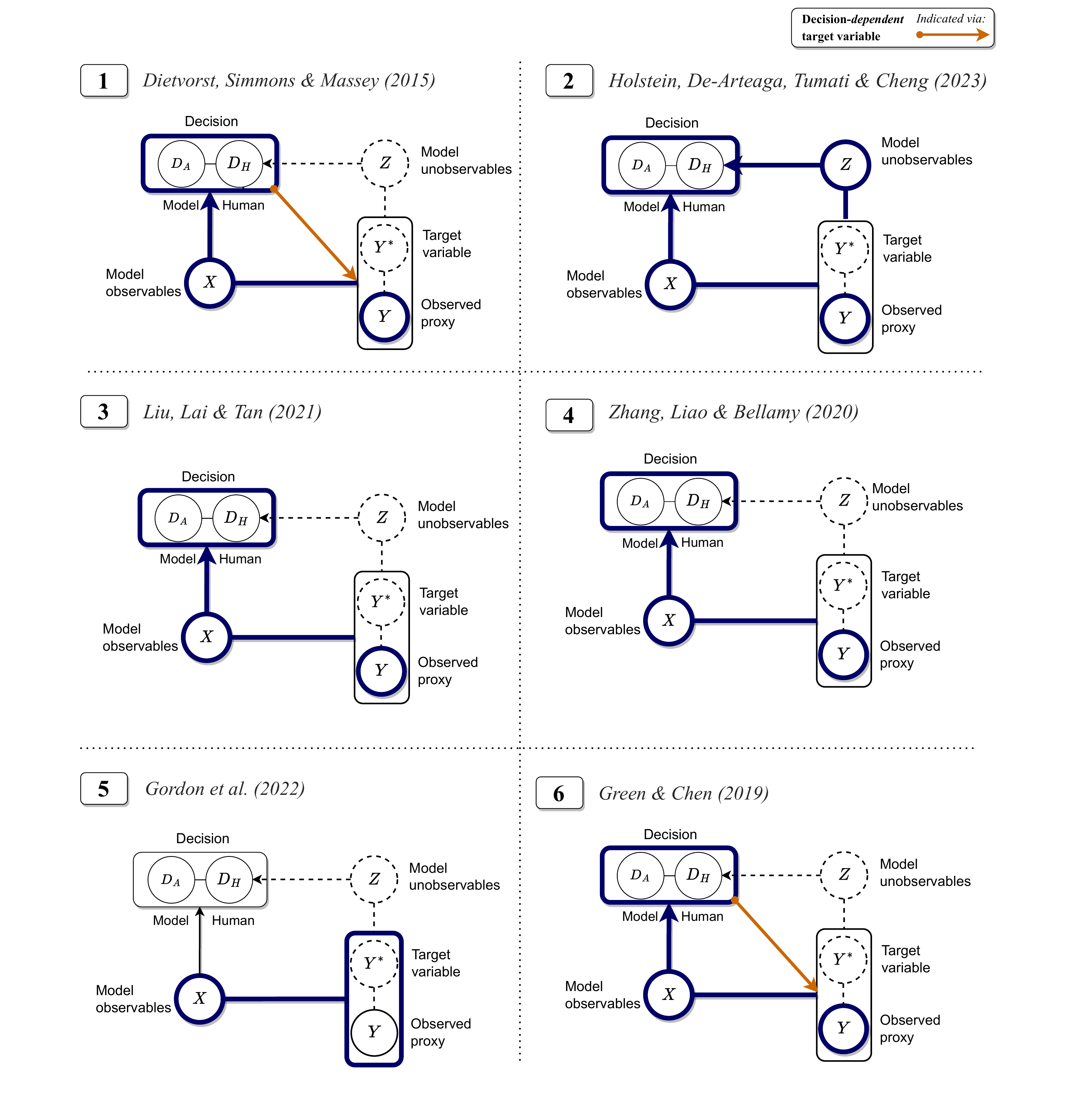}}
\caption{Causal diagrams for six of the studies included in our review of prior human-AI decision-making experiments discussed in $\S$ \ref{subsec:appendix_experimental_coding}. Sub-regions of focus in the study are shown with bold blue borders. The arrow connecting the $D$ node with target variables $Y$ and $Y^*$ is omitted in tasks falling under the decision-independent regime.}
\label{fig:study_supplement}
\end{figure*}

\clearpage

\lgedit{
\subsection{Extended review of prior experimental studies through the lens of our causal framework}\label{subsec:appendix_experimental_coding}

In this section, we provide a resource to help researchers examine factors related to target variable bias during the design and evaluation of experimental human-AI decision-making studies. We provide a detailed examination of several studies included in our review \citep{green2019principles, gordon2022jury, zhang2020effect, buccinca2021trust, holstein2023toward, dietvorst2015algorithm}. For each study, we identify (1) the \textbf{sub-region of focus}, and (2) the \textbf{ADS regime} used in the experimental evaluation.

\begin{itemize}
    \item The \textbf{sub-region of focus} describes the primary nodes and edges considered in the experimental design and evaluation of the work (e.g., regions shown in Figure \ref{fig:experimental_studies_dag}). This region can be determined by the description of the experimental design (i.e., conditions and RQs), task, and methods provided by the authors. For example, works often report the co-variates used to train a model ($X$), proxy label ($Y$), and experimental manipulation of focus in the study. For many studies included in our review, the experimental manipulation involves a modification to the joint decision region of our diagram ($D$) in the form of explanations \citep{zhang2020effect}, cognitive forcing functions \citep{buccinca2021trust}, model accuracy \citep{dietvorst2015algorithm}, or other behavioral interventions.

    \item The \textbf{ADS regime} describes the data generating process that gave rise to the dataset used to train the predictive model examined in the experimental evaluation. Our causal framework contains two specific ADS regimes: those with (1) decision-independent target variables and (2) decision-dependent target variables. In contrast to the sub-region of focus, the ADS regime is \textit{implicit} in the description of prior studies. This is because the majority of prior studies do not explicitly discuss factors related to outcome measurement error, unobservables, or treatment effects that may be relevant in the task design.
\end{itemize}

\subsubsection{Study 1: \citet{dietvorst2015algorithm}}

In this study, \citet{dietvorst2015algorithm} popularize the term \textit{algorithm aversion} by finding that ``participants more quickly lose confidence in algorithmic than human forecasters after seeing them make the same mistake.'' This study instructed participants to play the part of an MBA admissions officer by predicting the percentile the student would rank among their peers given application information such as undergraduate degree, GMAT scores, interview quality, essay quality, work experience, average salary, and parents’ education. The primary experimental manipulation studied whether participants would elect to use human judgement versus a statistical model given different information about their relative performance. We show the sub-region of focus and ADS regime in Figure \ref{fig:study_supplement}.1.

\begin{itemize}
    \item \textbf{Sub-region of focus: $D$, $X$, $Y$}. This study focuses on the sub-region with joint decisions $D$, co-variates $X$ and outcome proxies $Y$. We include co-variates ($X$) because the authors list an explicit set of features that are provided to both the human and the model. We include the joint human-model decision region ($D$) because the experimental treatment alters participant awareness of human and model performance differences. We include proxy labels ($Y$) because the authors describe an outcome variable of student ``success'', defined as an average of multiple performance measures (GPA, respect of fellow students, and prestige of employer upon graduation). We do not include $Y^*$ because the authors do not examine additional operationalizations of ``success'' or ``student performance'' that could be possible in this admissions setting. We do not include unobservables $Z$ because the authors do not examine other factors (e.g., student demeanor, personal connections) that might be available to an admissions officer but not a model. We do not include the edge connecting decisions $D$ and outcomes $Y$, $Y^*$ because the authors do not examine the impact of predictions and admissions decisions on downstream student performance. 

    \item \textbf{ADS Regime: decision-dependent target variable.} In this setting, the decisions of admissions offers determine which students are admitted to the graduate program, and, consequentially, which students have academic performance outcomes available. Recall from $\S$ \ref{sec:tasks_challenges} that this is a \textit{selective opportunity} setting because we only observe outcomes for students provided the enrollment opportunity. As a result, confounding and selection bias are relevant in this modeling task, in addition to outcome measurement error. In real-world deployments of predictive models for admissions decisions, unobservables and outcome measurement error may impact the ADS deployment due to private information available to a loan officer and alternate definitions of ``academic success'' or ``academic performance'' that may be relevant in this setting. \textbf{As a result, findings from this study may be most likely to generalize to other decision-dependent target variable tasks (e.g., financial loan approvals, commercial job hiring decisions, or pre-trial release decisions).} 
\end{itemize}

\subsubsection{Study 2: \citet{holstein2023toward}} 

In this study,the authors examine model unobservables as a potential source of complementary in an AI-assisted house price prediction task. Participants were shown a set of ``Facts and Features'' about homes (e.g., year built, type of heating, number of bathrooms, zoning classification) and asked to predict the house's sale price. These facts corresponded to tabular features available in the training data. Three of the eight features were removed during model training to introduce synthetic unobservables, and experimental conditions varied how participants were prompted to consider these unobservables during their decision-making. We show the sub-region of focus and ADS regime in Figure \ref{fig:study_supplement}.2.

\begin{itemize}
    \item \textbf{Sub-region of focus: $Z$, $D$, $X$, $Y$}. This study focuses on the sub-region with joint decisions $D$, model observables \textbf{$X$, model unobservables} $Z$, and outcome proxies $Y$. We include model observables ($X$) because the authors list an explicit set of features that were provided to both the human and the model. We include the joint human-model decision region ($D$) because the experimental treatment involved different participant prompts for considering unobservables during their decisions. We include the unobservables region ($Z$) because the authors explicitly omit predictive features from the model during training, but provide these to participants at decision time. We include the proxy label ($Y$) because the authors list a predictive outcome of house sale price. However, we do not include $Y^*$ because the authors do not examine other potential operationalizations of ``house worth'' possible in this task (e.g., the amount \textit{a participant would pay} for a house versus its actual market sale price). We do not include the edge connecting decisions $D$ and outcomes $Y$, $Y^*$ because the authors do not examine the impact of price predictions on downstream sales. 

    \item \textbf{ADS Regime: decision-independent target variable.} In this setting, the sale price predictions of participants does not impact downstream house sale prices. Therefore, we list this task as decision-independent target variable. While it is conceivable that loan officer, real estate agent, or online platform price predictions could impact house sale prices (e.g., Zestimates) in similar settings, this is \textbf{not} the case in this particular evaluation because there is not a decision being informed by the model that directly impacts observed prices. In particular, the historical data available for model training lists a full set of houses and their corresponding prices, with no prior human decisions/price predictions that might have impacted the price. Because observed prices are not connected to the prediction task in this study, we list this as decision-independent. Therefore, while outcome measurement error could be a concern in this setting due to differing notions of ``house quality'', selection bias, confounding, selective labels, and treatment effects are not a concern in this evaluation. \textbf{As a result, findings from this study may be most likely to generalize to other decision-independent target variable tasks (e.g., nutrient content prediction, forest cover prediction) and may be less likely to generalize to real-world predictive model deployments with decision-dependent outcomes.}
\end{itemize}

\subsubsection{Study 3: \citet{liu2021understanding}}

This study examines whether interactive explanations and out-of-distribution examples can foster human-AI complementary. Out-of-distribution examples refers to a setting in which the human-AI team makes decisions involving instances from a distribution that differs in composition from the model training dataset. The authors experimentally manipulate (1) sources of distribution shift and (2) presentation of interactive explanations. The authors conduct evaluations via recidivism prediction tasks (see Study 6 below) and an occupation classification task in which participants predict an individual's occupation given a written biography drawn from the BIOS dataset. We show the sub-region of focus and ADS regime for the occupation prediction task in Figure \ref{fig:study_supplement}.3. 

\begin{itemize}
    \item \textbf{Sub-region of focus:  $D$, $X$, $Y$}. This study focuses on the sub-region with joint decisions ($D$), model observables  ($X$), and outcome proxies ($Y$). We include model observables ($X$) because participants were shown a written biography about each person drawn from the BIOS dataset \citep{de2019bias}. We include decisions $D$ because the authors experimentally manipulate the explanation type and data distribution and examine impacts on human-AI decision quality. We include the proxy label ($Y$) because the authors list a prediction target involving the \textit{reported occupation} of an individual in the dataset (e.g., psychologist, physician, surgeon, teacher, and professor). We do not include $Y^*$ because the \textit{reported occupation} of individuals in the BIOS data can overlook reporting bias or multiple professions (e.g., physician and professor), which is not examined in the experimental manipulation. We do not include $Z$ because the study participants and model were both given access to the same biography information. We do not include the edge connecting decisions $D$ and outcomes $Y$, $Y^*$ because participant guesses do not influence the occupation of individuals in the BIOS data.  

    \item \textbf{ADS Regime: decision-independent target variable.} Because participant responses do not influence the occupations of individuals in the dataset, this is a task with decision-independent target variables. The authors do examine selection bias by modifying the distribution at run-time (e.g., out-of-distribution examples). \textbf{Therefore, this evaluation may generalize to ADS deployments in which models are subject to selection bias, but may generalize less readily to decision-dependent outcome tasks or those with pronounced outcome measurement error.} 
    
\end{itemize}

\subsubsection{Study 4: \citet{zhang2020effect}}

This study examines whether showing model confidence scores (probability estimates) and local explanations helps humans make more accurate decisions while using predictive models. This study also examines whether these decision-time interventions help humans better calibrate trust in the models predictions, defined as following recommendations more often when the model is more confident. To test this hypothesis, the authors trained a model to predict whether an individuals income would exceed $\$50K$ given tabular demographic and job information from the UCI Adult Data Set. We show the sub-region of focus and ADS regime in Figure \ref{fig:study_supplement}.4.

\begin{itemize}
    \item \textbf{Sub-region of focus:  $D$, $X$, $Y$}. This study focuses on the sub-region with joint decisions $D$, observables  $X$, and outcome proxies $Y$. We include model observables ($X$) because both the human and the model had access to the same set of 8 attributes about individuals while predicting their income. We include the proxy label ($Y$) because the authors list a target outcome involving whether an individual makes more or less than $\$50K$. We include joint human model decisions ($D$) because the experimental treatment involves different decision-time interventions shown to participants (i.e., model confidence scores or explanations). We do not include $Y^*$ because the authors do not examine sources of measurement error that can impact the reported income available in data. The authors leverage the UCI Adult dataset based on 1994 Census Data, which could be subject to various sources of reporting bias.  We do not include the edge connecting decisions $D$ and outcomes $Y$, $Y^*$ because participant guesses do not influence the income of individuals in the dataset.  

    \item \textbf{ADS Regime: decision-independent target variable.} Because participant responses do not influence the income of participants, this task includes decision-independent target variables. As a result, confounding, selection bias, intervention effects, and selective labels are not a concern in this task. \textbf{As a result, findings from this study may be most likely to generalize to other decision-independent target variable tasks (e.g., house price prediction, jellybean counting) and may be less likely to generalize to real-world predictive model deployments with decision-dependent outcomes.}
\end{itemize}

\subsubsection{Study 5: \citet{gordon2022jury}}

This work proposes a normative and technical framework called Jury Learning, which is intended to help practitioners \textit{``recognize and integrate annotator disagreement in the classifier pipeline \citep{gordon2022jury}.’’}  Under the proposed framework, model developers specify groups of users whose opinions should be considered during moderation decisions (i.e., juries), along with a relative weighting of each group. At inference time, a model predicts the annotations of each individual annotator, and a final decision is reached by combining predictions via the specified jury rule.\footnote{See \citep{gordon2022jury} for framework details not discussed in this summary, such as repeated sampling over several trials.} As part of the framework evaluation, the authors recruited online moderators from Discord, Twitch, and Reddit, and evaluated the diversity of annotator pools constructed via Jury Learning against a baseline of ``majority vote'' aggregation in a comment toxicity classification task. Thus, this study differs from those discussed above because the involvement of human subjects occurs at \textit{model development time} rather than at \textit{decision time}. Nevertheless, this toxicity classification task falls within our framework scope ($\S$ \ref{sec:scope}). We show the sub-region of focus and ADS regime in Figure \ref{fig:study_supplement}.5. 

\begin{itemize}
    \item \textbf{Sub-region of focus:  $X$, $Y$, $Y^*$}. We include model observables ($X$) because both the human and the model have access to the same set of information about comments. We include $Y$ and $Y^*$ because the study investigates how practitioners construct differing jury rules for mapping observed ratings from participants ($Y$) to the latent construct of ``toxicity'' being predicted by the model ($Y^*$). The $D$ region is not included in this study because the authors do not examine content moderation decisions or toxicity ratings at \textit{deployment time}. We omit unobservables $Z$ because the authors do not study how unobserved information could impact toxicity perceptions of annotators or the learned jury decisions. 

    \item \textbf{ADS Regime: decision-independent.} In this setting, the label targeted by toxicity classification models is determined by the subjective opinion of the annotator viewing the content. As a result, measurement error is relevant in this setting because the operationalization of ``toxicity'' targeted by the model depends on the identity of the user, the context in which the post is viewed, and the annotation protocol, among other factors. However, confounding, selection bias, selective labels, and treatment effects are not of concern in this setting because there is not a time dependency of decisions and outcomes. \textbf{Therefore, findings from this study may be most likely to generalize to other decision-independent target variable tasks (e.g., house price prediction, jellybean counting) and may be less likely to generalize to real-world predictive model deployments with decision-dependent outcomes.}

\end{itemize}

\subsubsection{Study 6: \citet{green2019principles}}

This work examines whether risk assessments improve the accuracy, fairness, and reliability of human decisions in financial lending and recidivism prediction tasks. The authors train risk assessments to predict re-arrest and loan default outcomes given tabular administrative data. The experimental conditions test several variations of the procedure for presenting risk assessment information to participants (e.g., no score, local explanation, immediate outcome feedback) before participants make the final decision. We show the sub-region of focus and ADS regime in Figure \ref{fig:study_supplement}.6. 

\begin{itemize}
    \item \textbf{Sub-region of focus:  $D$, $X$, $Y$}. We include the $D$ region because experimental conditions manipulate the joint human-model decision-making process. We include the $X$ region because participants were provided with a narrative profile containing factual content that coincides with the model training features (e.g., defendant age, applicant credit score). We include the proxy label region $Y$ because the authors list a target outcome consisting of failure to appear or re-arrest (recidivism) and loan default. We omit the target variable $Y^*$ because the authors do not examine sources of measurement error impacting recorded re-arrest and default outcomes (e.g., crimes that go unreported). We do not bold the arrow from $D$ to $Y$ and $Y^*$ because the authors do not examine how the historical decisions of judges or loan officers might influence the outcomes available for the applicant pool. 

    \item \textbf{ADS Regime: decision-dependent target variable.} Both experimental tasks included in this study involve a setting in which a model is trained on data from decisions made under an earlier decision-making policy. As a result, loan repayment is only observed among approved applicants, while re-arrest and failure to appear is only observed among released defendants. As a result, the model included in this experimental task is subject to selection bias, selective labels, confounding, intervention effects, and measurement error. \textbf{Therefore, findings from this study may be most likely to generalize to other decision-dependent target variable tasks.}

\end{itemize}
}